\documentclass[journal]{IEEEtran}
\usepackage{amsmath,amssymb,amsfonts}
\usepackage{algorithmic}
\usepackage{algorithm}
\usepackage{graphicx}
\usepackage{textcomp}
\usepackage{xcolor}
\usepackage{cite}
\usepackage{bm}
\usepackage{booktabs}   
\usepackage{tikz}
\usepackage{tikz-3dplot}
\usepackage{caption}
\usetikzlibrary{arrows.meta, calc}
\makeatletter
\newcommand{\captionof}[2]{\def\@captype{#1}\caption{#2}}
\makeatother
\usepackage{times}  
\usepackage{setspace}
\begin{document}
	
	\title{Joint Antenna Rotation and IRS Beamforming for Multi-User Uplink Communications}

	\author{Guoying~Zhang, 
		Qingqing Wu, 
		Ziyuan Zheng, 
		Qiaoyan Peng, 
		Yanze Zhu,
		Wen  Chen, 
		Penghui Huang
		\thanks{Guoying Zhang, Qingqing Wu, Ziyuan Zheng, Yanze Zhu, Wen Chen and Penghui Huang are with the Department of Electronic Engineering, Shanghai Jiao Tong University, Shanghai 200240, China (e-mail: \{gy\_zhang; qingqingwu;  zhengziyuan2024; yanzezhu; wenchen; huangpenghui\}@sjtu.edu.cn). Qiaoyan Peng is with the State Key	Laboratory of Internet of Things for Smart City, University of Macau, Macao	999078, China (e-mail: qiaoyan.peng@connect.um.edu.mo).} 

	}
	
	\maketitle
	
	\begin{abstract}
		Rotatable antenna (RA) enhances wireless coverage through directional gain steering, yet suffers from performance degradation under physical blockages. Intelligent reflecting surface (IRS) establishes reflective paths to bypass obstacles, but suffers from angular mismatch when deployed in the side-lobe region of base station (BS) antennas. To address this issue, we propose a new RA-enabled IRS-assisted multi-user uplink system, in which the BS antennas are capable of flexibly adjusting their 3D orientations to align their boresights with the IRS. We formulate a sum rate maximization problem by jointly optimizing the antenna 3D rotations, receive beamforming and IRS phase shifts. To tackle this non-convex problem, we propose an efficient alternating optimization (AO) algorithm. Specifically, we iteratively update the antenna rotations via projected gradient ascent (PGA), compute the receive beamforming via a closed-form solution, and optimize the IRS phase shifts via fractional programming (FP). Numerical results demonstrate that the proposed system yields significant performance gains over conventional fixed-antenna systems, especially under large angular misalignments.
	\end{abstract}
	
	
	\begin{IEEEkeywords}
		Intelligent reflecting surface (IRS), rotatable antenna (RA), alternating optimization (AO), uplink communication.
	\end{IEEEkeywords}
	
	\section{Introduction}
	\label{sec:intro}
	
	\IEEEPARstart{H}{igh}-frequency communications, such as millimeter-wave (mmWave) bands, offer abundant spectrum resources, yet signals suffer from severe penetration loss and blockage. 
	To compensate for path loss, high-gain directional antennas are essential~\cite{balanis2016antenna}. However, conventional base station (BS) antennas with fixed orientations limit the exploitation of spatial degrees of freedom (DoFs). To address this limitation, various flexible antenna architectures have emerged~\cite{zhu2024movable}. While advanced solutions such as the six-dimensional movable antenna (6DMA) optimize both position and rotation~\cite{Shao6DMA2024}, they incur high implementation complexity. As a cost-effective alternative, the rotatable antenna (RA) enhances coverage by adjusting only its orientation while keeping the position fixed, effectively avoiding complex position tracking~\cite{zheng2025RA}.
	
	Despite the beam-steering capability of RAs, reliable communication remains vulnerable to physical blockage, especially in dense urban environments. Intelligent reflecting surfaces (IRSs) provide a promising solution to this issue by establishing virtual line of sight (LoS) paths to bypass obstacles~\cite{wu2020towards}. However, large-scale IRSs required to compensate for the double-fading  effect caused by multiplicative path loss introduce geometric challenges \cite{Wu2025Intelligent}. Specifically, the Fraunhofer distance grows quadratically with aperture~\cite{cui2022channel}. This places the BS in the near-field region where each IRS reflecting element (RE) subtends a different angle~\cite{lu2024tutorial}. Moreover, due to the fixed orientation of conventional BS antennas, the IRS is often located far from the antenna boresight, leading to severe gain mismatch. 
	Such angular misalignment is particularly detrimental in multi-user systems, degrading both signal power and spatial discrimination for interference suppression. While position-flexible architectures such as movable intelligent surfaces~\cite{zheng2025MIS} and movable antenna-IRS integration~\cite{zheng2025MAIRS, Gao2025Integrating} can mitigate this misalignment to enhance system performance, they entail high costs. Conversely, existing research on the cost-effective RA-enabled IRS system is limited to single-user cases~\cite{cheng2022ris}. Thus, the joint design of BS-side RA and IRS for multi-user systems under near-field geometric constraints remains unexplored.
	
	Motivated by the above, we investigate an RA-enabled IRS-assisted multi-user uplink system, where each BS antenna independently adjusts its 3D orientation to compensate for the angular mismatch between the BS and the IRS. This joint design entails two key challenges: 1) the rotation variables are coupled with both direct and IRS-cascaded channels via distinct geometric dependencies, requiring coordinated optimization, and 2) the antenna gain function is nonlinear and non-differentiable at the half-space boundary, complicating gradient-based optimization. We formulate a sum rate maximization problem and develop an alternating optimization (AO) algorithm to address these challenges. {Specifically, the rotation angles are updated via projected gradient ascent (PGA) with subgradient handling for the nondifferentiability at the gain boundary, the receive beamforming matrix is then obtained in closed form under the minimum mean-square error (MMSE) criterion, and the IRS phase shifts are finally optimized using fractional programming (FP) combined with the Riemannian conjugate gradient (RCG).}
	Numerical results validate that the proposed scheme achieves substantial performance gains over the IRS-assisted system with fixed antenna orientations and effectively compensates for the angular mismatch-induced gain loss, demonstrating the practical benefits of exploiting rotational DoFs.
		

		\section{System Model and Problem Formulation}
		\label{sec:system_model}
		
		\begin{figure}[t!]
			\centering 
			\includegraphics[height=1.5in,width=2.5in]{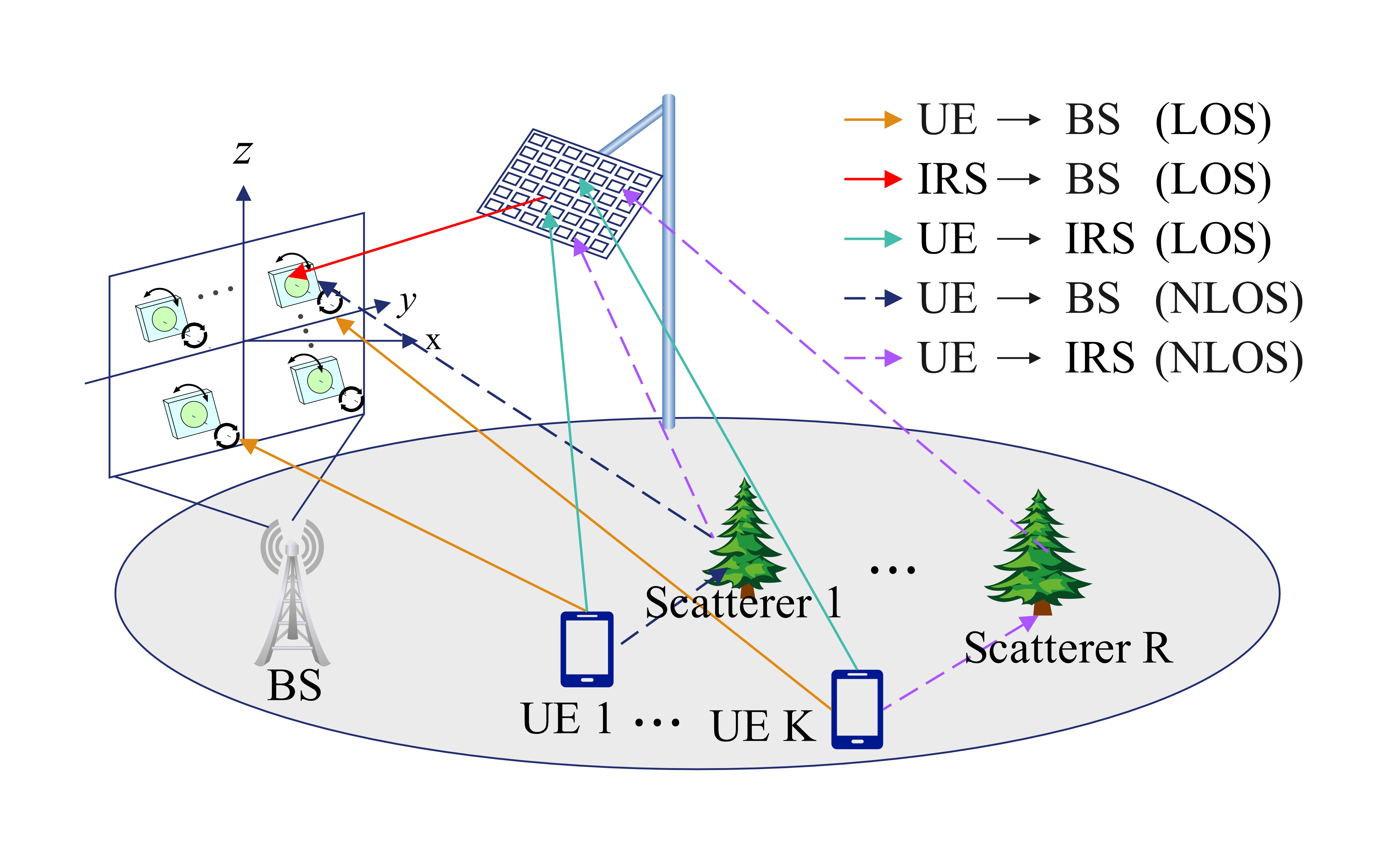}
			\caption{An IRS-assisted uplink communication system with RAs.}
			\label{system_model}
		\end{figure}
		
		As illustrated in Fig.~\ref{system_model}, we consider an uplink multi-user system where $K$ single-antenna users transmit signals to a BS equipped with a uniform planar array (UPA) of $M=M_x\times M_z$ RAs, indexed by $\mathcal{M} \triangleq \{1, \dots, M\}$, where $M_x$ and $M_z$ denote the number of antennas along the $x$-axis and $z$-axis, respectively. To bypass blockage in the direct links, an IRS comprising a UPA of $N=N_y\times N_z$ REs, indexed by $\mathcal{N} \triangleq \{1, \dots, N\}$, is deployed to assist the communication, with $N_y$ and $N_z$ representing the number of REs along the $y$-axis and $z$-axis, respectively.

		\subsection{RA-Enabled IRS-Assisted Uplink System Model}
		
		A global Cartesian coordinate system is established with the BS array center at the origin $\mathbf{b}_0=[0,0,0]^\top$. The BS array lies in the $x$--$z$ plane. The position of the $m$-th BS antenna is  given by  $\mathbf{b}_m=\mathbf{b}_0+\big[(m_x-\tfrac{M_x-1}{2})d_B,\,0,\,(m_z-\tfrac{M_z-1}{2})d_B\big]^\top$, 
		where $m_x\in\{0,\ldots,M_x-1\}$ and $m_z\in\{0,\ldots,M_z-1\}$ are the horizontal and vertical indices, respectively, and $d_B$ is the inter-antenna spacing.  
		The IRS center is located at $\mathbf{r}_0=[x_R,y_R,z_R]^\top$ with an outward normal vector $\mathbf{n}_{\mathrm{IRS}}$ oriented to enable effective reflection. The $N$ REs are uniformly distributed on the IRS panel with spacing $d_R$, and the position of the $n$-th RE is denoted by $\mathbf{r}_n=[x_n,y_n,z_n]^\top$. The $K$ users, indexed by $\mathcal{K} \triangleq \{1, \dots, K\}$, are located at $\mathbf{u}_k=[x_k,y_k,z_k]^\top$.
		
		Each RA can steer its boresight by mechanically tuning its elevation and azimuth angles\footnote{Each RA is driven by a compact two-axis actuator, avoiding the per-element position tracking and reconfigurable RF cabling required by movable antenna architectures. Independent RA orientations enable steering gain toward different directions, facilitating a balance between the IRS-reflected and direct links.}. {Let $\boldsymbol{\theta}_m\triangleq[\theta_m,\phi_m]^\top$ denote the deflection angle vector of the $m$-th antenna, and let $\boldsymbol{\Theta}\triangleq[\boldsymbol{\theta}_1,\boldsymbol{\theta}_2,\ldots,\boldsymbol{\theta}_M]$ denote the deflection angle matrix of all RAs}, where $\theta_m\in[0,\theta_{\max}]$ and $\phi_m\in[0,2\pi)$ are the elevation and azimuth angles of the $m$-th antenna relative to the positive $y$-axis, respectively. The parameter $\theta_{\max}$ represents the maximum mechanical rotation range. After rotation, the boresight direction of the $m$-th antenna becomes
		\begin{equation}
			\mathbf{f}_m({\boldsymbol{\theta}_m})=[\sin\theta_m\cos\phi_m,\,\cos\theta_m,\,\sin\theta_m\sin\phi_m]^\top.
			\label{eq:orientation}
		\end{equation}
		When $\theta_m=0$, the antenna points along its default boresight aligned with the positive $y$-axis, i.e., $\mathbf{f}_{\mathrm{ref}}=[0,1,0]^\top$. 
		Following the generic directional gain pattern~\cite{wu2025modeling}, the directional gain of the $m$-th antenna towards a signal arriving from direction $\mathbf{d}$ is modeled as
		\begin{equation}
			G(\epsilon)  = \begin{cases}
				G_{\max} \cos^{2p}(\epsilon), & \epsilon \in [0, \frac{\pi}{2}),  \\
				0, & \text{otherwise},
			\end{cases}
			\label{eq:antenna_gain}
		\end{equation}
		where $p \geq 1$ determines the directivity factor and $G_{\max} = 2(2p + 1)$ ensures power conservation. The effective off-boresight angle $\epsilon$ between the antenna boresight $\mathbf{f}_m$ and the signal arrival direction $\mathbf{d}$ is computed via $\cos(\epsilon) = \mathbf{f}_m({\boldsymbol{\theta}_m})^\top \mathbf{d}$, where $\mathbf{d}$ is the unit vector pointing from the BS antenna towards the signal source.  
		
		\subsection{Channel Model and Signal Model}
		\label{subsec:channel}
		
		We adopt a deterministic geometry-based channel model to capture the impact of antenna rotation. Let $\lambda$ denote the carrier wavelength and $D_{\mathrm{IRS}}$ represent the maximum aperture dimension (i.e., the diagonal length) of the IRS. The Fraunhofer distance is given by $d_F^{\mathrm{IRS}} = 2D_{\mathrm{IRS}}^2/\lambda$~\cite{balanis2016antenna}. In the considered scenario, the BS is assumed to be located within the near-field region of the IRS (i.e., $\|\mathbf{b}_0 - \mathbf{r}_0\| < d_F^{\mathrm{IRS}}$), necessitating a spherical-wave model for the IRS-BS channel. Conversely, the users are assumed to be located in the far-field region of the IRS (i.e., $\|\mathbf{u}_k - \mathbf{r}_0\| > d_F^{\mathrm{IRS}}$), allowing the user-side links to be modeled using planar waves. The effective channel $\mathbf{h}_k(\boldsymbol{\Theta},\boldsymbol{\Phi})\in\mathbb{C}^{M\times 1}$ for user $k$ is given by
		\begin{equation}
			\mathbf{h}_k(\boldsymbol{\Theta},\boldsymbol{\Phi})
			=\mathbf{h}_{d,k}(\boldsymbol{\Theta})
			+\mathbf{H}_{RB}(\boldsymbol{\Theta})\boldsymbol{\Phi}\mathbf{h}_{r,k},
			\label{eq:total_channel}
		\end{equation}
		where $\mathbf{h}_{d,k}(\boldsymbol{\Theta})\in\mathbb{C}^{M\times 1}$ is the direct channel, $\mathbf{H}_{RB}(\boldsymbol{\Theta})\in\mathbb{C}^{M\times N}$ is the IRS-BS channel, and $\mathbf{h}_{r,k}\in\mathbb{C}^{N\times 1}$ is the user-IRS channel, and  {$\boldsymbol{\Phi}\triangleq\mathrm{diag}\big(e^{j\varphi_1},\ldots,e^{j\varphi_N}\big)$} is the IRS phase-shift matrix, with	$\varphi_n\in[0,2\pi)$ denoting the phase shift of the $n$-th RE.
		
		Assuming users are in the far field of the BS, the direct channel is modeled as the superposition of one LoS path 
		($l = 0$) and $L$ non-line-of-sight (NLoS) scattering paths, i.e.,
		\begin{equation}
			\mathbf{h}_{d,k}(\boldsymbol{\Theta})
			={\sum\nolimits_{l=0}^{L}}\alpha_{k,l}\mathbf{D}_{k,l}(\boldsymbol{\Theta})\mathbf{a}_{\mathrm{BS}}(\mathbf{d}_{k,l}),\ k \in \mathcal{K}.
			\label{eq:direct_channel}
		\end{equation}
		The  complex path gain $\alpha_{k,l}$ $(l \in \{0,\ldots,L\})$  is given by
		\begin{equation}
			\small 
			\alpha_{k,l}=
			\begin{cases}
				\frac{\lambda}{4\pi d_{k,0}}e^{-j\frac{2\pi}{\lambda}d_{k,0}}, & l=0\ (\mathrm{LoS}),\\[6pt]
				\frac{\lambda\sqrt{\sigma_{l}}}{4\pi t_{k,l} d_{l,\mathrm{BS}}}e^{-j\frac{2\pi}{\lambda}(t_{k,l}+d_{l,\mathrm{BS}})+j\xi_{l}}, & l\geq 1\ (\mathrm{NLoS}),
			\end{cases}
			\label{eq:direct_path_gain}
		\end{equation}
		where $d_{k,0}=\|\mathbf{u}_k-\mathbf{b}_0\|$ is the LoS distance, $\sigma_{l} \in [0.1, 1]$~m$^2$ is the radar cross section (RCS) of scatterer $l$, $t_{k,l} = \|\mathbf{u}_k - \mathbf{s}_l\|$ denotes the distance from user $k$ to scatterer $l$, $d_{l,\mathrm{BS}} = \|\mathbf{s}_l - \mathbf{b}_0\|$ is the distance from scatterer $l$ to the BS, and $\xi_{l} \in [-\pi, \pi)$ is the random phase shift introduced by scatterer $l$.  The diagonal matrix {$\mathbf{D}_{k,l}(\boldsymbol{\Theta})\triangleq\mathrm{diag}\big(\sqrt{G(\epsilon_{1,k,l})},\ldots,\sqrt{G(\epsilon_{M,k,l})}\big)$} captures the antenna gain variation, where the effective off-boresight angle $\epsilon_{m,k,l}$ is determined by the projection $\cos(\epsilon_{m,k,l})=\mathbf{f}_m({\boldsymbol{\theta}_m})^\top\mathbf{d}_{k,l}$.  The BS array response vector $\mathbf{a}_{\mathrm{BS}}(\mathbf{d}_{k,l})$ has its $m$-th entry given by  $[\mathbf{a}_{\mathrm{BS}}(\mathbf{d}_{k,l})]_m=e^{j\frac{2\pi}{\lambda}\mathbf{d}_{k,l}^\top\mathbf{b}_m}$, where $\mathbf{d}_{k,l}$ is the unit direction vector of the $l$-th path, pointing from the BS to the user/scatterer.
		
		Similarly, the user-IRS channel is given by
		\begin{equation}
			\mathbf{h}_{r,k}
			={\sum\nolimits_{q=0}^{P}}\mu_{k,q}\sqrt{G_{\mathrm{IRS}}^{\mathrm{in}}(\vartheta_{k,q}^{\mathrm{in}})}\,\mathbf{a}_{\mathrm{IRS}}(\mathbf{d}_{k,q}^{\mathrm{in}}),\ k \in \mathcal{K}.
			\label{eq:ur_channel}
		\end{equation}
		Here, the path gain $\mu_{k,q}$ for $q \in \{0,\ldots,P\}$ follows the same form as \eqref{eq:direct_path_gain} with the LoS distance replaced by $d_{k,0}^{\mathrm{UR}}=\|\mathbf{u}_k-\mathbf{r}_0\|$. The term $G_{\mathrm{IRS}}^{\mathrm{in}}(\vartheta)$ represents the IRS element directional gain, defined as $G_{\mathrm{IRS}}^{\mathrm{in}}(\vartheta)=G_{\mathrm{IRS,max}}\cos^{2p_R}(\vartheta)$, where $G_{\mathrm{IRS,max}}=2(2p_R+1)$, with $p_R$ denoting the element directivity factor, and the effective cosine is $\cos(\vartheta_{k,q}^{\mathrm{in}})=\mathbf{n}_{\mathrm{IRS}}^\top\mathbf{d}_{k,q}^{\mathrm{in}}$. Finally, $\mathbf{a}_{\mathrm{IRS}}(\mathbf{d}_{k,q}^{\mathrm{in}})$ is the IRS array response with the $n$-th entry $[\mathbf{a}_{\mathrm{IRS}}(\mathbf{d}_{k,q}^{\mathrm{in}})]_n=e^{j\frac{2\pi}{\lambda}(\mathbf{d}_{k,q}^{\mathrm{in}})^\top(\mathbf{r}_n-\mathbf{r}_0)}$. Here, $\mathbf{d}_{k,q}^{\mathrm{in}}$ is the unit direction vector of the $q$-th path incident at the IRS, pointing from the IRS to the user/scatterer.
		
		Since the IRS--BS distance falls within the near-field region of the IRS, a spherical-wave model is adopted. The $(m,n)$-th entry of $\mathbf{H}_{RB}(\boldsymbol{\Theta})$ is given by
		\begin{equation}
			\small 
			[\mathbf{H}_{RB}(\boldsymbol{\Theta})]_{m,n}
			=\frac{\lambda}{4\pi d_{m,n}}e^{-j\frac{2\pi}{\lambda}d_{m,n}}
			\sqrt{G(\epsilon_{m,n}^{\mathrm{arr}})\,G_{\mathrm{IRS}}^{\mathrm{out}}(\vartheta_{n,m}^{\mathrm{out}})},
			\label{eq:rb_channel}
		\end{equation}
		where $d_{m,n}=\|\mathbf{b}_m-\mathbf{r}_n\|$ is the distance between the $m$-th RA and $n$-th RE.  The effective cosine of the angle-of-arrival (AoA) at the $m$-th BS antenna is $\cos(\epsilon_{m,n}^{\mathrm{arr}})=\mathbf{f}_m({\boldsymbol{\theta}_m})^\top\mathbf{d}_{n,m}^{\mathrm{arr}}$, 
		where $\mathbf{d}_{n,m}^{\mathrm{arr}}$ is the unit direction vector pointing from the BS antenna toward the RE, representing the signal arrival direction at the BS. The effective cosine of the angle-of-departure (AoD) from the $n$-th RE is $\cos(\vartheta_{n,m}^{\mathrm{out}})=\mathbf{n}_{\mathrm{IRS}}^\top\mathbf{d}_{n,m}^{\mathrm{out}}$, with $\mathbf{d}_{n,m}^{\mathrm{out}}=(\mathbf{b}_m-\mathbf{r}_n)/d_{m,n}$ and $\mathbf{d}_{n,m}^{\mathrm{arr}}=-\mathbf{d}_{n,m}^{\mathrm{out}}$. The term $G_{\mathrm{IRS}}^{\mathrm{out}}(\vartheta)$ represents the reflective gain of each RE, which follows the same form as $G_{\mathrm{IRS}}^{\mathrm{in}}(\vartheta)$.
		
		Based on the channel model in (\ref{eq:total_channel}), the received signal vector $\mathbf{y}\in\mathbb{C}^{M\times 1}$ at the BS is given by $\mathbf{y}={\sum\nolimits_{k=1}^{K}}\sqrt{P_k}\mathbf{h}_k s_k+\mathbf{n}$, where $P_k$ is the transmit power of user $k$, $s_k$ satisfies $\mathbb{E}[|s_k|^2]=1$, and $\mathbf{n}\sim\mathcal{CN}(\mathbf{0},\sigma^2\mathbf{I}_M)$ is additive white Gaussian noise (AWGN). We assume that the transmitted symbols from different users are statistically independent of each other and uncorrelated with the noise. The BS employs a linear receive beamforming matrix $\mathbf{W}=[\mathbf{w}_1,\ldots,\mathbf{w}_K]\in\mathbb{C}^{M\times K}$, and the estimated symbol for user $k$ is $\hat{s}_k=\mathbf{w}_k^{H}\mathbf{y}$. Accordingly, the signal-to-interference-plus-noise ratio (SINR) for decoding information from user $k$ is given by
		\begin{equation}
			\gamma_k(\mathbf{W},\boldsymbol{\Theta},\boldsymbol{\Phi})
			=\frac{P_k|\mathbf{w}_k^{H}\mathbf{h}_k(\boldsymbol{\Theta},\boldsymbol{\Phi})|^2}{{\sum\nolimits_{j\neq k}}P_j|\mathbf{w}_k^{H}\mathbf{h}_j(\boldsymbol{\Theta},\boldsymbol{\Phi})|^2+\sigma^2\|\mathbf{w}_k\|^2}.
			\label{eq:sinr}
		\end{equation}
		By defining {$\mathbf{v}\triangleq[e^{j\varphi_1},\ldots,e^{j\varphi_N}]^{\top}$} as the IRS phase-shift vector, such that $\boldsymbol{\Phi}=\mathrm{diag}(\mathbf{v})$, the effective channel in \eqref{eq:total_channel} can be rewritten as $\mathbf{h}_k(\boldsymbol{\Theta},\mathbf{v})=\mathbf{h}_{d,k}(\boldsymbol{\Theta})+\mathbf{Q}_k(\boldsymbol{\Theta})\mathbf{v}$, where $\mathbf{Q}_k(\boldsymbol{\Theta}) \triangleq \mathbf{H}_{RB}(\boldsymbol{\Theta})\mathrm{diag}(\mathbf{h}_{r,k})$.
		
		\subsection{Problem Formulation}
		
		We aim to maximize the achievable sum rate of all users by jointly optimizing the receive beamforming matrix $\mathbf{W}$, the RA {deflection angle matrix} $\boldsymbol{\Theta}$, and the IRS phase-shift vector $\mathbf{v}$. Accordingly, the optimization problem is formulated as
		\begin{subequations} \label{prob:P1}
			\begin{align}
				(\mathrm{P1}):\ \max_{\mathbf{W},\boldsymbol{\Theta},\mathbf{v}}\quad & {\sum\nolimits_{k=1}^{K}} \log_2(1+\gamma_k(\mathbf{W},\boldsymbol{\Theta},\mathbf{v})) \label{eq:P1_obj}\\
				\text{s.t.}\quad & 0\le \theta_m\le \theta_{\max},\ \forall m \in \mathcal{M}, \label{eq:P1_cons_a}\\
				& 0\le \phi_m < 2\pi,\ \forall m \in \mathcal{M}, \label{eq:P1_cons_b}\\
				& |v_n|=1,\ \forall n \in \mathcal{N}. \label{eq:P1_cons_c}
			\end{align}
		\end{subequations}
		Problem $(\mathrm{P1})$ is a non-convex optimization problem due to the following reasons: 1) the objective function \eqref{eq:P1_obj} is non-concave over the optimization variables, 2) the unit-modulus constraint \eqref{eq:P1_cons_c} defines a non-convex feasible set, and 3) the optimization variables $\mathbf{W}$, $\boldsymbol{\Theta}$, and $\mathbf{v}$ are intricately coupled in the effective channel $\mathbf{h}_k(\boldsymbol{\Theta},\mathbf{v})$. These factors together make it challenging to obtain the globally optimal solution.

		\section{Proposed Solution}
		\label{sec:algorithm}
		
		In this section, instead of jointly optimizing $\boldsymbol{\Theta}$,  $\mathbf{W}$ and $\mathbf{v}$ with high computational complexity, an AO algorithm is proposed to solve problem $(\mathrm{P1})$ by alternatively optimizing one variable with the others being fixed. 
		
		\subsection{BS Antenna {Deflection Angle} Optimization}
		\label{subsec:bs_rotation}

		With fixed $\mathbf{W}$ and $\mathbf{v}$, we optimize the BS {deflection angle matrix} $\boldsymbol{\Theta}$. The subproblem can be written as
		\begin{subequations}
			\begin{align}
				{(\mathrm{P2})}: \quad \max_{\boldsymbol{\Theta}} \quad & R(\boldsymbol{\Theta}) = {\sum\nolimits_{k=1}^{K}} \log_2(1+\gamma_k(\boldsymbol{\Theta})) \\
				\text{s.t.} \quad & 0 \le \theta_m \le \theta_{\max}, \quad \forall m \in \mathcal{M}, \\  & 0\le \phi_m < 2\pi, \quad \forall m \in \mathcal{M}.
			\end{align}
		\end{subequations}
		We solve  {$(\mathrm{P2})$} via  PGA. Applying the chain rule, the gradient with respect to $\theta_m$ is given by \eqref{eq:grad_full} at the top of next page, where $N_k \triangleq P_k |\mathbf{w}_k^H \mathbf{h}_k|^2$ and $D_k \triangleq {\sum\nolimits_{j \neq k}} P_j |\mathbf{w}_k^H \mathbf{h}_j|^2 + \sigma^2 \|\mathbf{w}_k\|^2$. The channel gradient $\frac{\partial \mathbf{h}_i}{\partial \theta_m}$ for any user $i \in \mathcal{K}$ is a sparse vector with only the $m$-th entry nonzero, given by \eqref{eq:dh_full} at the top of next page. The gradient $\frac{\partial R}{\partial \phi_m}$ follows the same structure with $\frac{\partial \mathbf{f}_m}{\partial \theta_m}$ replaced by $\frac{\partial \mathbf{f}_m}{\partial \phi_m}$.
		
		\begin{figure*}[!t]
			\normalsize
			\begin{equation} 
				\frac{\partial R}{\partial \theta_m} = {\sum\nolimits_{k=1}^{K}} \frac{1}{\ln 2 (1+\gamma_k)}  \frac{1}{D_k^2} \left[ D_k  2P_k \operatorname{Re}\left\{ (\mathbf{h}_k^H \mathbf{w}_k) \mathbf{w}_k^H \frac{\partial \mathbf{h}_k}{\partial \theta_m} \right\} - N_k {\sum\nolimits_{j \neq k}} 2P_j \operatorname{Re}\left\{ (\mathbf{h}_j^H \mathbf{w}_k) \mathbf{w}_k^H \frac{\partial \mathbf{h}_j}{\partial \theta_m} \right\} \right].
				\label{eq:grad_full}
			\end{equation}
			\begin{equation} 
				\left[\frac{\partial \mathbf{h}_i}{\partial \theta_m}\right]_m = {\sum\nolimits_{l=0}^{L}} \alpha_{i,l} \frac{\partial \sqrt{G(\epsilon_{m,i,l})}}{\partial \theta_m} [\mathbf{a}_{\mathrm{BS}}(\mathbf{d}_{i,l})]_m + {\sum\nolimits_{n=1}^{N}} \frac{\lambda e^{-j\frac{2\pi}{\lambda}d_{m,n}}}{4\pi d_{m,n}} \sqrt{G_{\mathrm{IRS}}^{\mathrm{out}}(\vartheta_{n,m}^{\mathrm{out}})} \frac{\partial \sqrt{G(\epsilon_{m,n}^{\mathrm{arr}})}}{\partial \theta_m} [\mathbf{h}_{r,i}]_n v_n, \quad \forall i \in \mathcal{ K }.
				\label{eq:dh_full}
			\end{equation}
			\hrulefill
		\end{figure*}
		
		The main challenge is that $G(\epsilon)$ is non-differentiable when $\epsilon = \pi/2$. Expressing $G$ as a function of $c = \cos(\epsilon) = \mathbf{f}_m^\top \mathbf{d}$, we have $\tilde{G}(c) = G_{\max} \max(c, 0)^{2p}$, whose subgradient is
		\begin{equation}
			\tilde{\partial}_c \tilde{G}(c) = \begin{cases}
				2p G_{\max} c^{2p-1}, & \text{if } c > 0, \\
				0, & \text{otherwise}.
			\end{cases}
			\label{eq:subgrad_G}
		\end{equation}
		For the boresight direction in \eqref{eq:orientation}, the partial derivatives are $\frac{\partial \mathbf{f}_m}{\partial \theta_m} = [\cos\theta_m \cos\phi_m, -\sin\theta_m, \cos\theta_m \sin\phi_m]^\top$ and $\frac{\partial \mathbf{f}_m}{\partial \phi_m} = [-\sin\theta_m \sin\phi_m, 0, \sin\theta_m \cos\phi_m]^\top$. For $c = \mathbf{f}_m^\top \mathbf{d} > 0$, applying the chain rule yields
		\begin{equation}
			\frac{\partial \sqrt{G}}{\partial \theta_m} = \sqrt{G_{\max}}  p  c^{p-1}  \left(\frac{\partial \mathbf{f}_m}{\partial \theta_m}\right)^\top \mathbf{d},
			\label{eq:grad_theta}
		\end{equation}
		and similarly for $\phi_m$; when $c \leq 0$, the gradient is zero.  
		
		Based on the above discussion, the update rule for $\boldsymbol{\Theta}$ is 
		\begin{equation}
			\boldsymbol{\Theta}^{(t+1)} = \mathcal{P}_{\mathcal{C}}\left(\boldsymbol{\Theta}^{(t)} + \eta^{(t)} \nabla_{\boldsymbol{\Theta}} R\right),
			\label{eq:grad_update}
		\end{equation}
		where $\mathcal{P}_{\mathcal{C}}(\cdot)$ projects onto the feasible set  $\mathcal{C} = \{(\theta_m,\phi_m): 0 \le \theta_m \le \theta_{\max}, 0 \le \phi_m < 2\pi, \forall m\}$ and $\eta^{(t)}$ is determined by backtracking line search with Armijo condition to ensure monotonic improvement.
		
		\subsection{Receive Beamforming Optimization}
		\label{subsec:beamforming}
		
		For any given {deflection angles} $\boldsymbol{\Theta}$ and phase shifts $\mathbf{v}$, the effective channel $\mathbf{h}_k(\boldsymbol{\Theta}, \mathbf{v})$ becomes a constant vector $\mathbf{h}_k$. 
		It is well known that the MMSE receiver maximizes the sum rate in multi-user uplink systems~\cite{tse2005fundamentals}.  Up to a scaling factor that does not affect the SINR, the MMSE solution is given by
		\begin{equation}
			\mathbf{w}_k^{\star} = \left( {\sum\nolimits_{j=1}^K} P_j \mathbf{h}_j \mathbf{h}_j^H + \sigma^2 \mathbf{I}_M \right)^{-1} \mathbf{h}_k, \forall k \in \mathcal{ K}.
			\label{eq:mmse}
		\end{equation}
		
		\subsection{IRS Phase Shift Optimization}
		\label{subsec:irs_phase}
		
		With fixed $\mathbf{W}$ and $\boldsymbol{\Theta}$, we optimize the IRS phase shifts $\mathbf{v}$. Since $\mathbf{v}$ is subject to the unit-modulus constraint, it lies on a complex circle manifold $\mathcal{A} = \{ \mathbf{x} \in \mathbb{C}^N : |x_n|=1, \ \forall n \}$. The subproblem for optimizing $\mathbf{v}$ can be expressed as
		\begin{subequations}
			\begin{align}
				{(\mathrm{P3})}: \ \max_{\mathbf{v}} \quad & {\sum\nolimits_{k=1}^{K}} \log_2\left(1+\frac{S_k(\mathbf{v})}{I_k(\mathbf{v})}\right) \label{eq:P2_obj} \\
				\text{s.t.} \quad & |v_n| = 1, \quad \forall n  \in \mathcal{N}, \label{eq:P2_cons}
			\end{align}
		\end{subequations}
		where $S_k(\mathbf{v}) = P_k |\mathbf{w}_k^H (\mathbf{h}_{d,k} + \mathbf{Q}_k \mathbf{v})|^2$ and $I_k(\mathbf{v}) = {\sum\nolimits_{j \neq k}} P_j |\mathbf{w}_k^H (\mathbf{h}_{d,j} + \mathbf{Q}_j \mathbf{v})|^2 + \sigma^2 \|\mathbf{w}_k\|^2$ denote the desired signal power and interference-plus-noise power, respectively. To handle the fractional objective and unit-modulus constraint, we employ the FP technique based on the quadratic transform~\cite{shen2018fractional}. By introducing auxiliary variables $\boldsymbol{\beta} = [\beta_1, \ldots, \beta_K]^{\top}$, problem  {$(\mathrm{P3})$} is reformulated as
		\begin{subequations}
			\begin{align}
				{(\mathrm{P3'})}: \ \max_{\mathbf{v}, \boldsymbol{\beta}} \quad & {\sum\nolimits_{k=1}^{K}} g_k(\mathbf{v}, \beta_k) \label{eq:fp_problem} \\
				\text{s.t.} \quad & |v_n| = 1, \quad \forall \in \mathcal{N}, \label{eq:fp_cons}
			\end{align}
		\end{subequations}
		where {$g_k(\mathbf{v}, \beta_k) \triangleq \log_2(1+2\operatorname{Re}\{\beta_k^* a_k(\mathbf{v})\} - |\beta_k|^2 I_k(\mathbf{v}))$} with {$a_k(\mathbf{v}) \triangleq \sqrt{P_k} \mathbf{w}_k^H (\mathbf{h}_{d,k} + \mathbf{Q}_k \mathbf{v})$}. For fixed $\mathbf{v}$, the optimal auxiliary variable is $\beta_k^\star =  a_k^*(\mathbf{v}) /I_k(\mathbf{v})$. For fixed $\boldsymbol{\beta}$, the objective  involves quadratic terms in $\mathbf{v}$, which we optimize via the RCG algorithm on the manifold $\mathcal{A}$~\cite{shen2018fractional}.
		
		\begin{algorithm}[t]
			\caption{Proposed AO Algorithm for Solving (P1)}
			\label{alg:ao}
			\begin{algorithmic}[1]
				\STATE \textbf{Input:} Initial $\boldsymbol{\Theta}^{(0)}$, $\mathbf{v}^{(0)}$, tolerance $ \varepsilon $, max iterations $T$.
				\STATE \textbf{Output:} Optimized $\mathbf{W}^\star$, $\boldsymbol{\Theta}^\star$, $\mathbf{v}^\star$.
				\STATE  Set $R^{(0)} = -\infty$. 
				\FOR{$t = 1, 2, \ldots, T$}
				\STATE  {Update $\boldsymbol{\Theta}^{(t)}$ by solving (P2) via PGA in \eqref{eq:grad_update}.}
				\STATE  {Update $\mathbf{W}^{(t)}$ via MMSE in \eqref{eq:mmse}.}
				\STATE  {Update $\mathbf{v}^{(t)}$ by solving (P3') via FP and RCG.}
				\STATE  Compute $R^{(t)} = {\sum\nolimits_{k=1}^{K}} \log_2(1+\gamma_k^{(t)})$. 
				\IF{$|R^{(t)} - R^{(t-1)}| <  \varepsilon $}
				\STATE \textbf{break}
				\ENDIF
				\ENDFOR
				\STATE \textbf{return} $\mathbf{W}^{(t)}$, $\boldsymbol{\Theta}^{(t)}$, $\mathbf{v}^{(t)}$.
			\end{algorithmic}
		\end{algorithm}
		
		\subsection{Overall Algorithm and Complexity}
		\label{subsec:convergence}
		
		The proposed AO algorithm is summarized in Algorithm~\ref{alg:ao}. Since the objective value is non-decreasing and upper-bounded, the algorithm is guaranteed to converge.  {In each iteration, the complexity is dominated by the updates of $\boldsymbol{\Theta}$ and $\mathbf{v}$. The update of $\mathbf{W}$ involves matrix inversion with complexity $\mathcal{O}(M^3)$. The complexity of updating $\boldsymbol{\Theta}$ via PGA is $\mathcal{O}(I_{\mathrm{GA}} M K)$, while updating $\mathbf{v}$ via RCG has a complexity of $\mathcal{O}(I_{\mathrm{RCG}} N)$. Therefore, the total complexity per iteration is $\mathcal{O}(M^3 + I_{\mathrm{GA}} M K + I_{\mathrm{RCG}} N)$, where $I_{\mathrm{GA}}$ and $I_{\mathrm{RCG}}$ denote the number of iterations for gradient ascent and RCG, respectively.}

		\section{Simulation Results}

		\begin{figure*}[t]
			\centering
			\captionsetup{justification=centering} 
			
			\begin{minipage}[t]{0.3 \linewidth}
				\centering
				\includegraphics[width=0.85\linewidth, height=1.7in]{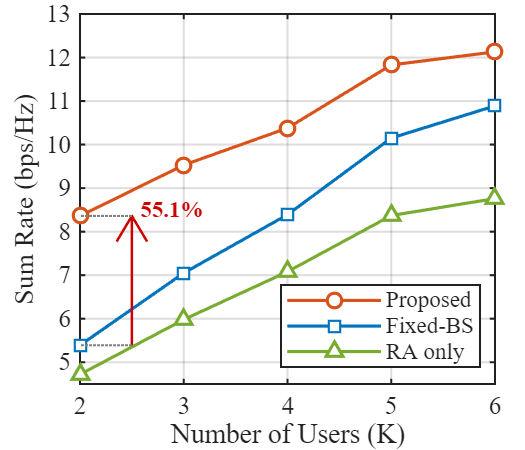}
				\captionof{figure}{Sum rate vs. users $K$.}
				\label{fig:fig2_vs_users}
			\end{minipage}
			\hfill
			\begin{minipage}[t]{0.3  \linewidth}
				\centering
				\includegraphics[width=0.85\linewidth, height=1.7in]{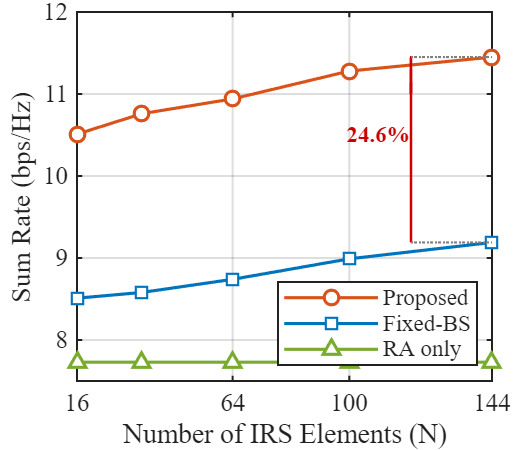}
				\captionof{figure}{Sum rate vs. IRS elements $N$.}
				\label{fig:fig3_vs_irs1}
			\end{minipage}
			\hfill
			\begin{minipage}[t]{0.3  \linewidth}
				\centering
				\includegraphics[width=0.85\linewidth, height=1.7in]{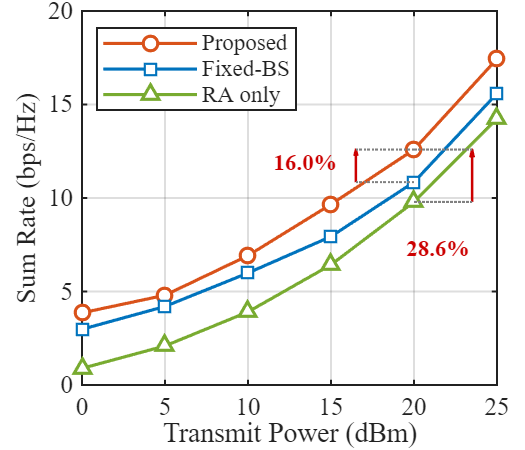}
				\captionof{figure}{Sum rate vs. power $P_k$.}
				\label{fig:fig4_vs_power}
			\end{minipage}
			\vspace{-2em}
		\end{figure*}

		{\begin{table}[t]
				\centering
				\caption{Simulation Parameters}
				\label{tab:params}
				\setlength{\tabcolsep}{3pt} 
				\resizebox{\columnwidth}{!}{%
					\begin{tabular}{lc||lc}
						\toprule
						\textbf{Parameter} & \textbf{Value} & \textbf{Parameter} & \textbf{Value} \\
						\midrule
						Carrier frequency $f_c$ & $28$ GHz & Noise power $\sigma^2$ & $-72$ dBm \\
						Transmit power $P_k$ & $17$ dBm & Direct link atten. & $20$ dB \\
						BS antennas $M$ & $16$ ($4\times4$) & IRS elements $N$ & $64$ ($8\times8$) \\
						Users $K$ & $4$ & Max rotation $\theta_{\max}$ & $60^\circ$ \\
						Directivity $p$ / $p_R$ & $2$ / $1.5$ & NLoS paths $L$ / $P$ & $3$ / $2$ \\
						BS location & $[0,0,0]^\top$ m & IRS location & $[1.5,2,2]^\top$ m \\
						User region ($z=0$) & $[-3,3]\times[8,15]$ m & Element spacing & $\lambda/2$ \\
						\bottomrule
					\end{tabular}
				} 
		\end{table}}
		Simulation parameters are listed in Table~\ref{tab:params}, where the direct link attenuation models obstacle-induced blockage. The IRS aperture is configured to place the BS within the near-field region: the BS--IRS distance $\|\mathbf{b}_0-\mathbf{r}_0\| \approx 3.2$~m is strictly smaller than the Fraunhofer distance $d_F^{\mathrm{IRS}} = 2D^2/\lambda \approx 5.4$~m, where $D = 12\sqrt{2}$~cm denotes the IRS diagonal length. For comparison, we consider the following benchmark schemes:
		\begin{itemize}
			\item \textit{Fixed-BS+IRS}: The BS antenna orientations are fixed along the positive $y$-axis 
			, while the IRS phase shifts and receive beamforming are optimized.
			\item \textit{RA-only}: The antenna rotations and receive beamforming are jointly optimized without IRS assistance. 
		\end{itemize}

		Fig.~\ref{fig:fig2_vs_users} depicts the achievable sum rate versus the number of users $K$. 
		The proposed RA+IRS scheme consistently achieves the highest sum rate, outperforming Fixed-BS+IRS by {$1.24$--$2.97$~bps/Hz} and RA-only by {$3.30$--$3.64$~bps/Hz}. Notably, RA-only exhibits the slowest growth since blocked users rely heavily on weak NLoS links and cannot benefit from the IRS path regardless of antenna orientation. The gain over Fixed-BS+IRS peaks at {$55.1\%$ ($K=2$)} and narrows to {$11.4\%$ ($K=6$)}. This trend reflects the transition from a noise-limited to an interference-limited regime: at a small $K$, RA steering yields substantial channel gains, while at a large $K$, MMSE interference suppression becomes dominant. Nevertheless, the proposed scheme maintains a robust advantage over RA-only, demonstrating the indispensable role of the IRS in extending coverage { across diverse blockage patterns}.
		
		Fig.~\ref{fig:fig3_vs_irs1} evaluates the sum rate versus the number of IRS elements $N$.
		The RA-only scheme maintains a constant {$7.73$~bps/Hz} regardless of $N$, confirming that the IRS-reflected path is essential for users experiencing direct-link blockage.
		Both IRS-assisted schemes exhibit sum rate improvements as $N$ increases.
		However, the growth is sublinear rather than the squared power gain predicted by ideal passive beamforming theory.
		Specifically, {at $N=144$, the proposed RA+IRS scheme achieves a sum rate of $11.45$~bps/Hz, significantly outperforming the Fixed-BS+IRS scheme ($9.19$~bps/Hz). This corresponds to a performance gain of $24.6\%$, as explicitly illustrated in Fig.~\ref{fig:fig3_vs_irs1}.}
		This sublinear scaling is attributed to the interference-limited regime and the path loss variations inherent to near-field propagation where phase variations are significant. 
		Notably, the performance gap between the proposed scheme and Fixed-BS+IRS expands from {$2.00$~bps/Hz at $N=16$ to $2.26$~bps/Hz at $N=144$}, indicating that the RA's orientation flexibility becomes increasingly valuable for larger IRSs.
		This is because the RA can dynamically adjust its boresight to align with the effective angular spread of the near-field IRS aperture, whereas the fixed-orientation antenna suffers from growing angular mismatch as the IRS subtends a larger solid angle, thereby failing to effectively harvest the energy reflected from the peripheral IRS elements. 
		
		Fig.~\ref{fig:fig4_vs_power} illustrates the achievable sum rate versus the transmit power $P_k$. As expected, the sum rate with all considered schemes increases as the transmit power increases. The proposed RA+IRS scheme consistently achieves the highest performance, improving from {$3.86$~bps/Hz at $P_k = 0$~dBm to $22.80$~bps/Hz at $P_k = 30$~dBm}. Specifically, at $P_k = 20$~dBm, the proposed scheme achieves {$16.0\%$} and {$28.6\%$} performance improvement over the Fixed-BS+IRS and RA-only schemes, respectively. {This performance advantage is particularly pronounced in the low-to-moderate power regime, where the RA's orientation optimization provides substantial channel gains { especially when blockage is severe}. As transmit power increases, the relative gain diminishes since inter-user interference becomes the dominant limiting factor.}
		
		\section{Conclusion}
		\label{sec:conclusion}
		
		This paper investigated an RA-enabled IRS-assisted multi-user uplink system incorporating a hybrid near-field/far-field channel model to address the angular mismatch caused by constrained IRS deployment locations. An AO algorithm was developed to jointly optimize the {antenna {deflection angles}, receive beamforming, and IRS phase shifts}. Specifically, the antenna {deflection angles are} {updated} via PGA with subgradient treatment for the non-differentiable gain boundary, the receive beamforming is {computed} via a closed-form solution, and the IRS phase shifts are  {optimized} through FP with RCG. Simulation results demonstrated that the proposed scheme achieves up to $11.5\%$ sum rate improvement over Fixed-BS+IRS at moderate transmit power and maintains a consistent $2$--$4.7$~bps/Hz advantage over RA-only across all operating conditions. Furthermore, the performance gap between the proposed scheme and Fixed-BS+IRS expands as the IRS aperture increases, highlighting the critical role of antenna {deflection angle} optimization in compensating for the angular spread of large near-field IRS arrays. Collectively, these results substantiate that the rotational degree of freedom effectively mitigates the geometric mismatch loss inherent in fixed arrays, ensuring that the high-dimensional passive beamforming gain offered by the IRS is fully realized.

		\begin{spacing}{0.9}
			\bibliographystyle{IEEEtran}
			\bibliography{BSReference}
		\end{spacing}

	\end{document}